\documentclass[preprint,12pt]{elsarticle}

\usepackage{amssymb}
\usepackage{amsmath}
\usepackage{textcomp}
\usepackage{makecell}
\usepackage{bm}
\usepackage{siunitx}
\usepackage{threeparttable}
\usepackage{placeins}
\usepackage{booktabs}
\usepackage{xurl}

\usepackage[table]{xcolor} 
\usepackage{array}

\def\nuc#1#2{\relax\ifmmode{}^{#1}{\protect\text{#2}}\else${}^{#1}$#2\fi}

\journal{Nuclear Instruments and Methods A}

\makeatletter
\def\ps@pprintTitle{%
   \let\@oddhead\@empty
   \let\@evenhead\@empty
   \let\@oddfoot\@empty
   \let\@evenfoot\@oddfoot
}
\makeatother
\usepackage[colorlinks=true,linkcolor=blue,citecolor=blue,urlcolor=blue]{hyperref}

\begin{document}
\begin{frontmatter}

\title{Novel High-Radiopurity Doped Amorphous Silicon Resistors for Low-Background Detectors}

\author{






A.~Anker, P.C.~Rowson\textsuperscript{*}, K.~Skarpaas\\
\textit{SLAC National Accelerator Laboratory} \\[0.8em]

S.~Tsitrin \\
\textit{Marvell Berkeley Nanofabrication Laboratory}\\[0.8em]

I.J.~Arnquist, L.~Kenneth S.~Horkley, L.~Pagani, T.D.~Schlieder\\
\textit{Pacific Northwest National Laboratory}\\[0.8em]

E.~van Bruggen, P.~Kachru, A.~Pocar, N.~Yazbek\\
\textit{University of Massachusetts}\\[0.8em]

\begingroup
  \renewcommand\thefootnote{*}%
  \footnotetext{Corresponding author: rowson@slac.stanford.edu}%
\endgroup

}
\begin{abstract}

We present the results of a study of lightly doped amorphous silicon used as a resistive medium for high-radiopurity resistors in nuclear and particle physics research instrumentation.  Prototypes are produced for a Time Projection Chamber design for the nEXO neutrinoless double-beta decay search experiment that meet requirements for ultra-high radiopurity, good mechanical, cryogenic and high voltage performance, as well as useful vacuum ultraviolet (VUV) reflectivity.  Further study is warranted to refine production methods and to confirm that the technology used here is useful for more general applications.

\end{abstract}
\end{frontmatter}

\section{Introduction}
The work described here was motivated by the need for high-radiopurity components for the Time Projection Chamber (TPC) detector designed for the nEXO neutrinoless double beta decay ($0 \nu \beta \beta$) search experiment \cite{nEXO2021}. In a TPC the positions and magnitudes of collected ionization charge and scintillation light produced by high energy particles and photons are reconstructed.  The proposed design would deploy dual-purpose structural-member and voltage-divider resistors for the high-voltage field cage that is required for the charge drift region of the TPC \cite{nexoHV}. The nEXO TPC is immersed in liquid xenon (LXe) that serves as the signal source and detection medium.  The extraordinarily low levels of background gamma radiation required for this rare nuclear decay search (the expected half-life is $>10^{26}$ years) place severe constraints on all construction materials used for the apparatus, including electronic components.  In particular, uranium and thorium contaminants are required to be below the ppt level.

The TPC requires a uniformly graded electric field from cathode to anode and for this reason several voltage-divider resistors separating field-shaping surfaces are typically placed between the electrodes. In a previous version of our experiment, thick-film resistors were commercially produced on a sapphire substrate and successfully deployed in the EXO-200 TPC \cite{EXO200}, but this technology would lead to unacceptably high radioactive backgrounds in the 40 times (in total LXe mass) larger and more sensitive nEXO detector. In addition, no vendor could be found that was willing to develop alternative thick-film processes for our application.

 We decided to investigate thin-film options and contacted two industrial firms \cite{Anritsu,ViaMEMS} that suggested that we investigate low pressure chemical vapor deposition (LPCVD) \cite{LPCVD} polysilicon, but these firms were not positioned to initiate the required R\&D.  It is known that chemical vapor deposition of silicon is inherently an ultra-high purity process which is supported by the cleanroom procedures typically maintained at microfabrication facilities.  Our own previous experience in the radioassay of materials for our experimental $0 \nu \beta \beta$ research had revealed that fused silica is available with sufficiently high radiopurity. Sapphire as a substrate for resistor fabrication was also expected to be suitable, although this was unconfirmed at the time.  We decided to initiate our investigations in house at the Stanford Nanofabrication Facility (SNF) \cite{SNF}.  We obtained high purity Suprasil 300 fused silica tubing from Heraeus Inc. of the approximate dimensions needed (6 mm OD, 4 mm ID).  

 At SNF we produced our first set of prototypes with undoped ``intrinsic'' silicon of 1.4~$\mu$m thickness with satisfactory mechanical results but the resistance of these parts proved to be far too high, even at this large layer thickness.  Our work continued at the Marvell Berkeley Nanofabrication Laboratory (MBNL) \cite{MBNL} to examine whether doping could be used to produce Si at our desired resistivity.  
 
 All of these activities will be described in what follows. First the design motivation and requirements are outlined for the spacer resistors within nEXO. Then the initial run at SNF is overviewed along with the cold testing of the undoped Si resistors at SLAC. Next the fabrication and testing of the doped Si resistors at MBNL will be described in detail. In the latter sections, the radiopurity and UV reflectivity of these resistors are quantified.

 \section{nEXO Motivated Design}

A number of features of the proposed nEXO LXe TPC design contribute to a set of constraints for voltage-divider resistors to be used in the HV field cage. The relevant features are:

\begin{itemize}
    \item The TPC is operated at LXe temperature (165~K).
    \item The resistors have high radiopurity (U, Th at ppt levels).
    \item The resistor chain material does not introduce electronegative emanation. 
    \item The resistors double as structural members in the tensioned field cage and must be mechanically robust under compressive loading .
    \item The resistors are cylindrical tubes with a resistive coated outer surface and with an inner diameter to accommodate end-to-end tensioning rods for the field cage.
    \item Both ends of the resistors are metallized for contact with field-shaping rings.
    \item The resistors must stably withstand 2 kV potential difference end-to-end (at a cathode voltage of -50~kV the potential difference across each resistor is $\sim$1~kV.) 
    \item Reflectivity of the resistive coating in the vacuum ultra violet (VUV) 175~nm wavelength is useful.
\end{itemize} 

\noindent These dual-purpose resistors, hereafter referred to as 
spacer/resistors, are required to have the properties listed above except for the last one listed. The desirability of reflectivity at 175 nm arises due to the need to collect as much of the expected VUV scintillation light produced in the LXe, a need that depends on the total solid angle reflected by these parts. 

Our design studies concluded that the desired resistance per resistor in our voltage divider at our operating temperature of 165~K should be in the range of 0.1 G$\Omega$ to 10.0 G$\Omega$, due to constraints on the maximum current draw from the HV power supply and the maximum acceptable RC discharge time constant (The estimated capacitance of the completed TPC field cage is 3~nF, corresponding to a few minutes for a maximum total resistance of the voltage divider of approximately 100 GOhm).  Acceptable field uniformity is achieved if the variation in the set of resistors within a divider chain is less than 20\%.

\begin{figure}[htbp]
     \centering
     \includegraphics[width=0.5\linewidth]{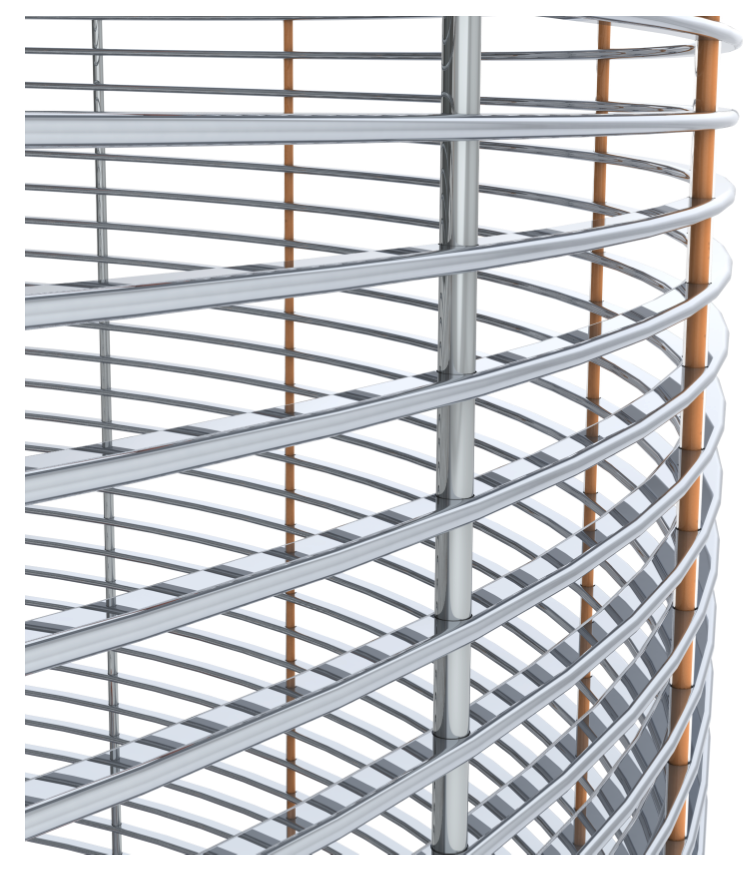}
     \caption{Rendering of the nEXO field cage.  The spacer/resistors have a silvery finish, while the ordinary fused silica or sapphire spacers are transparent and reveal the tensioning rods (here shown with a contrasting amber color) that pass through them.  The copper field shaping rings (a total of 57) are aluminized.}
     \label{fig:fieldcage}
 \end{figure}

 Figure~\ref{fig:fieldcage} shows the conceptual layout of the TPC field cage, with the field-shaping rings separated by 24 dielectric spacers through which anode to cathode tensioning rods pass through holes in the rings and through the inner diameter of the spacers, to compress the cage.  Some of the spacers (nominally 6 out of 24) are dual-purpose spacer/resistors that make up the voltage divider. For the nominal nEXO field cage, these parts have length, outer diameter (OD) and inner diameter (ID) of approximately 17 mm, 6 mm and 4 mm respectively. The spacer/resistors have a resistive coating on their outer surface except at the ends which are metalized to make conductive contact with each field-shaping ring, thereby forming the voltage divider circuit.  The absence of connectors/fasteners within the resistor chain and the insetting of the resistor ends from the edges of the field-shaping rings avoids the associated sharp points and high electric field regions, thereby reducing HV performance risk. The material used for these resistors (silicon thin film, fused silica or sapphire substrate, and the thin metallization) are not expected to produce chemical emanation.

\section{Amorphous and Polycrystalline Silicon: Initial R\&D at SNF}
 
In our initial studies at SNF, fused silica substrates of the approximate dimensions required for the TPC spacer/resistors were made from 6 mm OD, 4 mm ID Heraeus Suprasil 300 \cite{Heraeus} tubing, cut into 17 mm lengths at SLAC. It was determined early on that a small chamfer of the outer circumference of each part was useful to improve metallization integrity. The parts were placed in the LPCVD furnace six at a time in trial runs using custom fixturing, with pure silane ($\rm SiH_4$) gas used for Si coating at four different temperatures: 530\textdegree C, 580\textdegree  C, 590\textdegree  C and 620\textdegree  C. In all cases LPCVD runs produced Si layers of approximately 1.4 $\mu$m thickness, chosen due to the large resistivity expected for undoped intrinsic Si. Table~\ref{tab:SNFruns} shows the results for our four prototype runs at SNF. It was observed that the two runs at lower temperature (530\textdegree  C and 580\textdegree  C) produced a coating with a mirror-like finish and good reproducibility in measured resistivity part-to-part. In comparison, the high temperature runs had a grey, patchy appearance, a lower resistivity by more than an order of magnitude, and larger variation in resistance among the parts. 

As required for our design concept, our first undoped amorphous Si prototypes made at SNF included metallization at both ends.  E-beam evaporation was used to metallize the parts with an adhesion layer of 10 nm Ti followed by 100 nm of Pd. Palladium was chosen due to high expected radiopurity and low brittleness compared to platinum.  This metallization procedure was followed for all prototypes at SNF and MBNL. A qualitative change occurs at a furnace temperature between 580\textdegree C and 590\textdegree C, manifested by the order of magnitude jump in mean resistance, the increased variability of the parts as well as the changed appearance of the Si surface.

We were advised \cite{Kamins} that apparently a transition temperature existed between 580\textdegree C and 590\textdegree C between the amorphous and polycrystalline states of Si, with the amorphous state produced below 590\textdegree C, and that a precise transition temperature value was not found in the literature. While we did not confirm the crystallinity of these different coatings by independent means, our observations convinced us that (plausibly) amorphous silicon (aSi) was more appropriate for our application.


\begin{table}[ht]
    \centering
    \caption{The room temperature resistance for the SNF prototype runs with LPCVD undoped Si at various furnace temperatures, each run consisting of six spacer/resistors.  The uncertainties are the standard deviation (SD) of the set of resistors tested at each temperature.} 
    \quad
    \setlength{\tabcolsep}{4pt} 
    \renewcommand{\arraystretch}{1.2}
    \begin{tabular}{|c|ccc|}
    \rowcolor{gray!15}
    \hline
      & \bf{LPCVD temp} & \bf{Resistance (room temp)} & \bf{Appearance} \\
    \hline
    \bf{Run 1} & \SI{580}{\celsius} & $5.63 \pm 0.78 \text{G} \Omega$ & mirror, uniform \\

    \bf{Run 2} & \SI{530}{\celsius} & $10.91 \pm 0.19 \text{G} \Omega$ & mirror, uniform \\

    \bf{Run 3} & \SI{620}{\celsius} & $0.33 \pm 0.08 \text{G} \Omega$, SD \textgreater20\% & matte, nonuniform \\

    \bf{Run 4} & \SI{590}{\celsius} & $0.20 \pm 0.11 \text{G} \Omega$, SD \textgreater50\% & matte, nonuniform \\
    \hline
    \end{tabular}
    \label{tab:SNFruns}
\end{table}

\begin{figure}[htbp]
     \centering
     \includegraphics[width=0.9\linewidth]{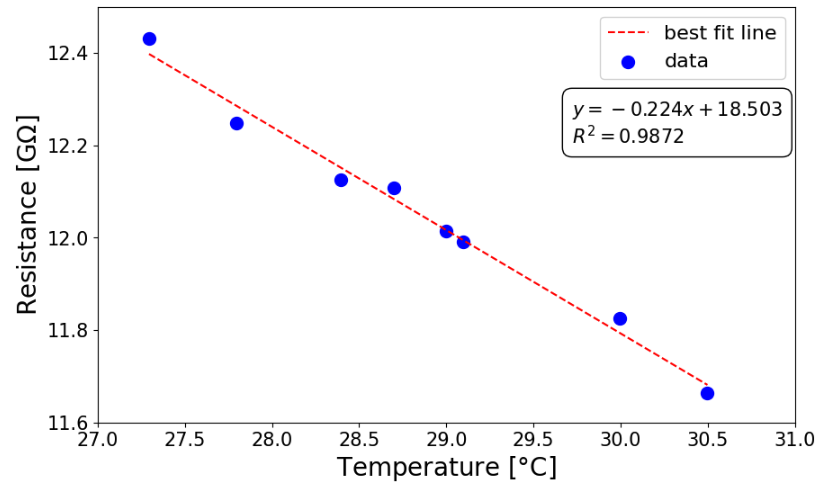}
     \caption{Illustration of the thermal coefficient of resistivity (TCR) at near room temperature of a single SNF-fabricated undoped aSi prototype (selected from Run 2, 530\textdegree C deposition). The fitted TCR corresponds to $-1.9 \times 10^{-2} \rm { ~per~degree ~K}$ at 12 GOhm.}
     \label{fig:roomtempTCR}
 \end{figure}

Note that the target resistance for our spacer/resistor was anticipated to be approximately 1 G$\Omega$ at 165~K. Even for the large Si thickness we used, which was close to a practical maximum, it was apparent that lower resistivity was needed. Another discovery was the observation of a very large and negative thermal coefficient of resistivity (TCR) at room temperature as shown in Figure~\ref{fig:roomtempTCR}, which made the need for lower resistivity unavoidable.  This behavior is typical for intrinsic silicon, where it is expected that the magnitude of the TCR increases exponentially at lower temperatures.  

\section{Experimental setup at SLAC}
\label{sec:SLACSETUP}

\begin{figure}[htbp]
     \centering
     \includegraphics[width=0.6\linewidth]{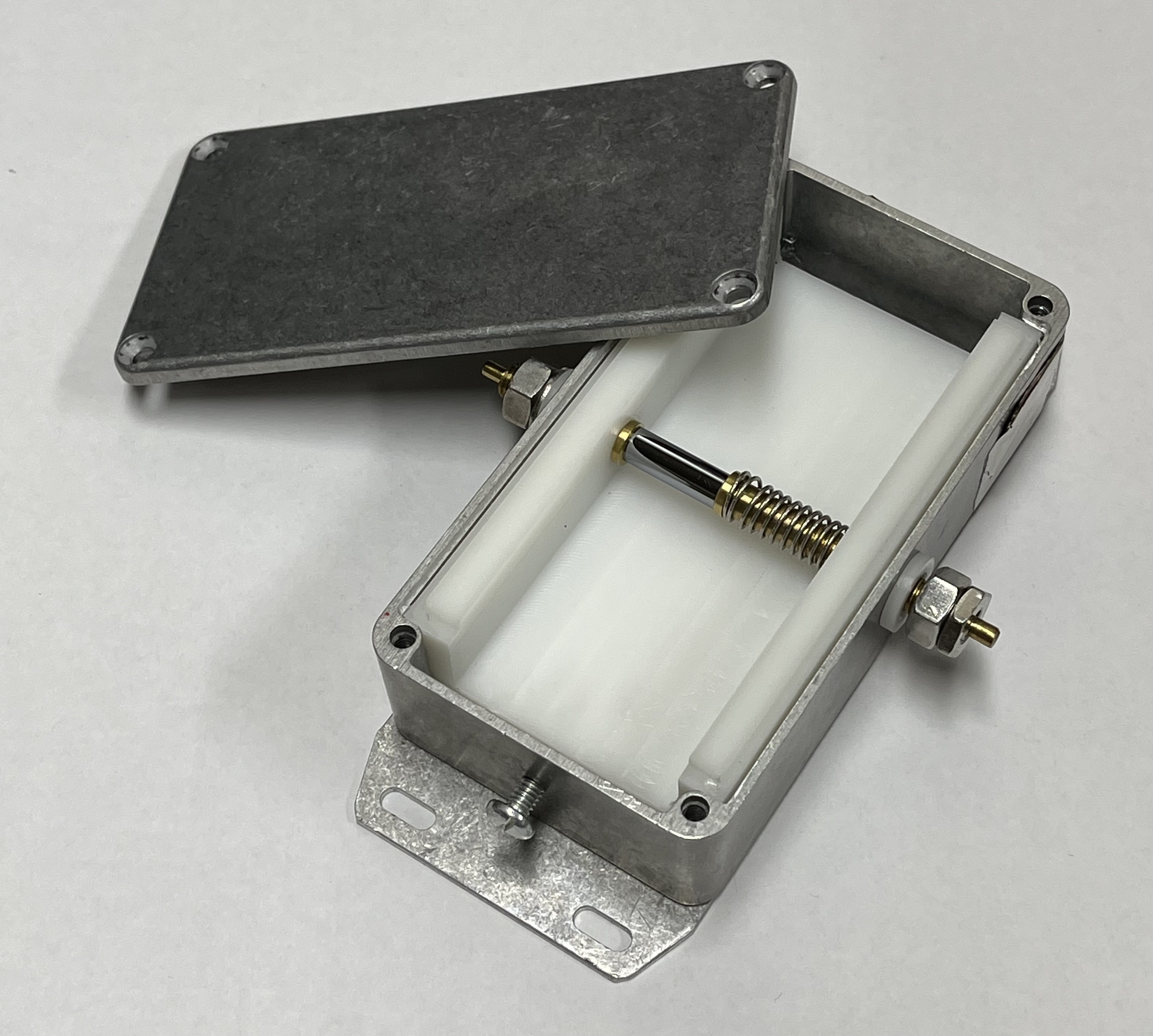}
     \caption{Metal testing box for spacer/resistors with custom spring fixturing to make contact under compression.}
     \label{fig:testing_box}
 \end{figure}

It was realized early on in our investigations that cryogenic testing of our prototypes was essential.  To do this, we used an Environmental Chamber made by Sun Electronic Systems model EC12 \cite{ECref}, capable of providing a controlled enclosure at our desired operating temperature of 165 K.  A small aluminum box, see Figure~\ref{fig:testing_box}, was modified to provide fixturing and electrical shielding for the thermal testing of the prototypes in the EC while a Keithley 6517B \cite{Keithley}  electrometer/high resistance meter monitored resistance.  A measurement of an amorphous silicon prototype that had a resistance of about 10 G$\Omega$ at room temperature measured nearly 2 T$\Omega$ at 165~K. In order to reach our target resistivity, the silicon needed to be lightly doped. 

We were advised at SNF that the doping equipment available at the time was unlikely to reliably produce low dopant concentrations and were pointed to colleagues at Berkeley as an alternative.  We then approached the Director of the Marvell Berkeley Nanofabrication Laboratory (MBNL) for assistance with our R\&D program.

\section{Phosphorous Doped Silicon R\&D at MBNL}

\begin{figure}[htbp]
     \centering
     \includegraphics[width=0.7\linewidth]{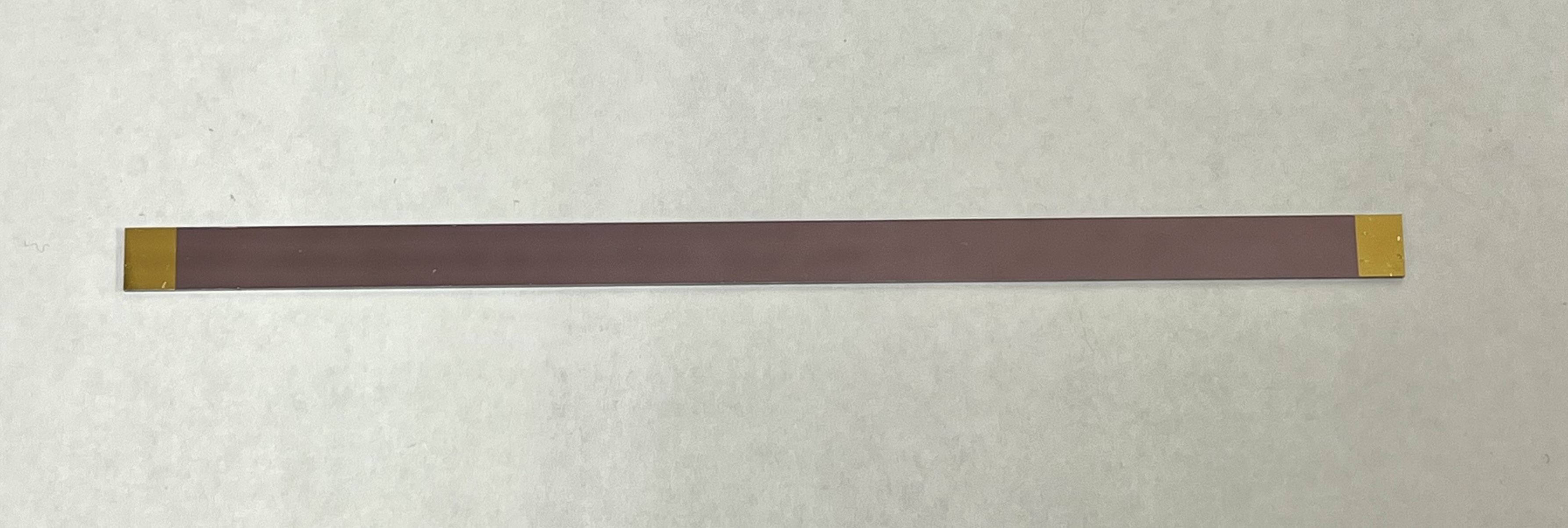}
     \caption{Segment of a diced wafer (a ``strip'') with gold plating on the ends and dimensions 10 cm x 0.5 cm.}
     \label{fig:strip}
 \end{figure}

At MBNL we devised a program using phosphorous doping with their Tystar model Mini-Tytan 3800 Horizontal LPCVD furnace \cite{Tystar}. Their system included a recently installed lower range gas flow controller (GFC) from Pivotal Systems \cite{Pivotal}, capable of flow rates as low as 0.2 standard cubic centimeters per minute (sccm).  A CHA Industries model Solution E-beam Evaporator \cite{CHA}  was used for metallization.  We used 150 mm silicon and fused silica wafers for trial resistivity runs, where a 200 sccm silane ($\rm SiH_4$) flow rate was supplemented by varying flow rates of a dopant gas : a 50/50 mixture of silane and phosphine ($\rm PH_3$).  The furnace temperature was kept at 550\textdegree C for all runs to give us the best chance of depositing amorphous film at a reasonable deposition rate. Following the LPCVD process, metal contact strips (10~nm ~Ti undercoating with a 100 nm ~top Au layer) were evaporated perpendicular to the flat at both ends of each wafer. The wafers were then diced into 10~cm long x 0.5~cm wide rectangles (see Figure~ \ref{fig:strip}) which would fit in the testing fixture used for spacer/resistor prototypes. These diced planar resistors served as prototypes in resistivity testing to determine the needed doping recipe.

Note that the poor radiopurity of gold relative to the platinum group metals (an observation made for the EXO-200 experiment) is not relevant for these preliminary strip tests.  

Table~\ref{tab:dopingtest} shows our results for P doping-level experiments.  The flow rates of the phosphine/silane dopant mixture varied between 0.4 and 1.0 sccm, with a pure silane flow rate of 200 sccm.  Silicon wafers, due to their mirror surface, were used for film thickness determination by spectroscopic ellipsometry. For the sheet resistance results, fused silica wafers were used, as the insulating substrate prevented the measurement confusion possible with Si wafers. From the surface resistance, deposited layer thickness, and part dimensions we deduce the projected resistance of a spacer/resistor as well as the resistivity of our doped aSi.  Reliable data were obtained for a subset of these tests at dopant mixture flow rates of 0.7 and 1.0 sccm. Lower dopant levels at $\le$~0.5~sccm corresponded to sheet resistance levels beyond our dynamic range.

\begin{table}[htbp]
    \centering
    \caption{Results of doping concentration tests.  Shown are the results for test wafers produced with two different silane/phosphine 50/50 dopant mixture flow rates (pure silane flow at 200 sccm in all cases).  For each doping concentration the following results are given: The sheet resistance, the aSi layer thickness, the resistivity, and the projected spacer/resistor resistance at 295~K for the standard dimensions of 17 mm length, 6 mm OD, 4 mm ID produced with the same doped aSi properties.}
    \quad
    \setlength{\tabcolsep}{5pt} 
    \renewcommand{\arraystretch}{1.2}
    \begin{tabular}{|m{2.5cm}|m{2.cm}|m{2cm}|m{1.9cm}|m{3.2cm}|}
    \rowcolor{gray!15}
       \hline
        \multicolumn{1}{|>{\centering\arraybackslash}m{2.5cm}}{\bfseries dopant flow [sccm]}
        & \multicolumn{1}{|>{\centering\arraybackslash}m{2cm}}{\bfseries sR [k$\Omega$/sq]}
        & \multicolumn{1}{|>{\centering\arraybackslash}m{2cm}}{\bfseries thickness [nm]}
        & \multicolumn{1}{|>{\centering\arraybackslash}m{2cm}}{$\boldsymbol{\rho}$ [$\Omega$ cm]}
        & \multicolumn{1}{|>{\centering\arraybackslash}m{3.2cm}|}{\bfseries projected spacer/resistor (295~K) [M$\Omega$]} \\
        \hline
        1.0 & $483 \pm 8$ & $281 \pm 3$ & $13.6 \pm 0.3$ & $0.436\pm 0.01$ \\
        0.7 & $3075 \pm 200$ & $330 \pm 10$ & $101 \pm 7.5$ & $2.77 \pm 0.2$ \\
        \hline
    \end{tabular}
\label{tab:dopingtest}
\end{table}

The procedures we use to characterize our produced aSi parts are common to all the results we obtained at MBNL and are summarized as follows : 

\noindent We used standard nano-fabrication methods to determine the properties of our aSi films.   Sheet resistance is provided by four-point resistance mapping instruments \cite{CDEresmap}, spectroscopic ellipsometry and reflectometry \cite{Ellips2,Nanoduv}  as well as a mechanical technique known as profilometry \cite{Dektak} give us the film thickness.  The precision of the data for sheet resistance and thickness of each film sample arises from instrumental effects that are estimated and reported by each device on a measurement-by-measurement basis. The uncertainties in the physical results are propagated into the calculation of derived quantities of interest such as the film resistivity or the expected resistance of a fabricated spacer/resistor.   If necessary, ``witness wafers", both silicon (with an oxide layer) or silica, are used as control substrates and are included in the furnace during the aSi CVD coating process.  For sheet resistance and ellipsometry measurements, multiple points on each sample or witness wafer surface are measured and the observed variation contributes to the reported measurement uncertainty. 

\subsection{Low Temperature testing}

After measuring the resistance of the wafer strips at room temperature, the testing was redone in the EC at 165~K as described in section \ref{sec:SLACSETUP}. The total resistivity change from 295~K to 165~K due to TCR effects, a multiplicative factor we designate as $\Delta \rm{TCR}$, determines whether or not a particular doping level would produce a spacer/resistor within our target range of 0.1~-~10.0~G$\Omega$ at the design operating temperature.  
Is was observed that a process using 0.7 sccm flow of the silane/phosphine mixture with a LPCVD layer thickness of about 300~nm achieved the required resistance range. The sheet resistance (measured using the 4-point contact method at MBNL) and aSi thickness (measured using ellipsometry, also at MBNL) are used to calculate the aSi resistivity. This resistivity is then used to predict the resistance of our standard spacer/resistor of dimensions (17 mm length, 6 mm OD, 4 mm ID) at room temperature, and using $\Delta \rm{TCR}$, also at 165~K.   

\subsection{Fabrication of spacer/resistors}

\begin{figure}[htbp]
     \centering
     \includegraphics[width=0.5\linewidth]{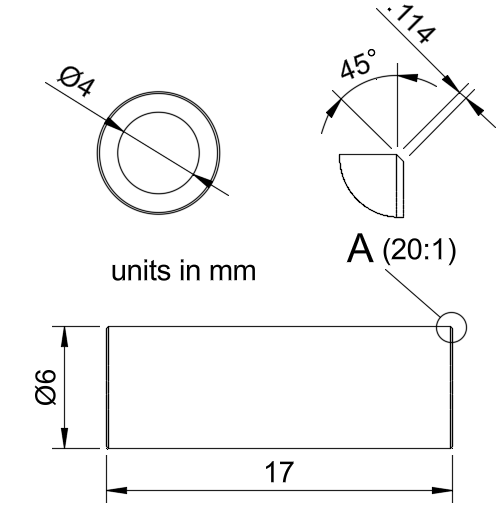}
     \caption{Fabrication drawing for the MBNL spacer/resistors.  The chamfer at the ends, done by hand for the SNF parts, is designed to improve the integrity of the metallization coating. }
     \label{fig:SR_drawing}
 \end{figure}

We identified a commercial shop \cite{Ferrites} for the preparation of 50 Heraeus Suprasil 300 tubing ``blanks'' of our standard dimensions, to the specifications shown in Figure~\ref{fig:SR_drawing}. 

We then produced a set of spacer/resistors using the same P-doping procedure to achieve equivalent P concentration and resistivity. A ``pilot'' run used transversely arranged 2 mm diameter fused silica rods and longitudinally wafer segments to suspend five fused silica spacer blanks threaded onto the rods. Following metallization, the pilot-run spacer/resistors were resistance tested and three were sent out for radiopurity testing discussed in the next section. 

Note that a direct optical ellipsometry measurement of aSi layer thickness used for wafers could not be made for the non-flat spacer/resistor surface.  Control of aSi layer thickness is done by placing a ``witness'' wafer in the furnace for each spacer/resistor run, which is then checked optically.

We observed that the produced parts showed smaller resistance than expected for aSi coating of the outer surface of the part; however, our fixturing does not prevent aSi coating of the inner ``bore'' of each spacer as the support rods are 2 mm smaller in diameter than the inner bore (4 mm).  In fact, the results showed that the inner bore is apparently coated to the same thickness as the outer surface, and given the 6~mm to 4~mm radius ratio of these surfaces, which constitute resistors in parallel, the expected resistance is 3/5 the size of the outer surface alone. To a good approximation, this factor of 0.6 is what we observe.  There is no obvious disadvantage that we are aware of in having outer and inner resistive coating on these parts, but if necessary a more elaborate fixturing scheme could be used to ``mask out'' the inner bore.

\begin{table}[t]
    \centering
    \caption{Results of the pilot spacer/resistor run for 0.7 sccm silane/phosphine 50/50 dopant mixture flow rate (pure silane flow at 200 sccm). Shown are the sheet resistance, the aSi layer thickness, and the resistivity as provided by the witness wafer, as well as the measured room temperature part resistance.  The projected spacer/resistor resistance at 295 K for the standard dimensions of 17 mm length, 6 mm OD, 4 mm ID produced with the same doped aSi properties are given for comparison.}
    \quad
    \setlength{\tabcolsep}{5pt} 
    \renewcommand{\arraystretch}{1.2}
    \rowcolors{1}{gray!15}{white}

    \begin{tabular}{|m{2.5cm}|m{2cm}|m{2cm}|m{2cm}|m{3.2cm}|}
        \hline
        \multicolumn{1}{|>{\centering\arraybackslash}m{2.5cm}}{\bfseries dopant flow [sccm]}
        & \multicolumn{1}{|>{\centering\arraybackslash}m{2cm}}{\bfseries sR [k$\Omega$/sq]}
        & \multicolumn{1}{|>{\centering\arraybackslash}m{2cm}}{\bfseries thickness [nm]}
        & \multicolumn{1}{|>{\centering\arraybackslash}m{2cm}}{$\boldsymbol{\rho}$ [\bfseries $\Omega$ cm]}
        & \multicolumn{1}{|>{\centering\arraybackslash}m{3.2cm}|}{\bfseries projected spacer/resistor (295 K) [M$\Omega$]} \\
        \hline
        0.7 & $3764 \pm 209$ & $278 \pm 31$ & $105 \pm 13$ & $3.39 \pm 0.57$ \\
        \hline
    \end{tabular}
\label{tab:pilotrunwitness}
\end{table}

Witness wafers, fused silica (for sheet resistance data) and silicon (for thickness data), were included in the furnace for the pilot run. The results for these wafers are given in Table~\ref{tab:pilotrunwitness}. Of the six spacer/resistors produced, three were not used for radiopurity testing or other purposes and were found to have the average characteristics given in Table~\ref{tab:pilot}.  

\begin{table}[ht]
    \centering
    \caption{The pilot run with 3 spacer resistors from MBNL.} \quad
    
    \renewcommand{\arraystretch}{1.2}
    \begin{tabular}{|l|c|}
    \rowcolor{gray!15}
    \hline
     \bfseries Properties & \bfseries Pilot Run \\
     \hline
     Mean Resistance (295~K) [M$\Omega$] & $2.02 \pm 0.11$ \\
     Standard dev (295~K) [M$\Omega$] & 0.19 \\
     Mean Resistance (165~K) [M$\Omega$] & $610 \pm 63$  \\
     Standard dev (165~K) [M$\Omega$] & 109 \\
     Mean $\Delta$TCR & $302 \pm 35$ \\ 
     \hline
    \end{tabular}
    \label{tab:pilot}
\end{table}

\begin{figure}[t]
     \centering
     \includegraphics[width=0.6\linewidth]{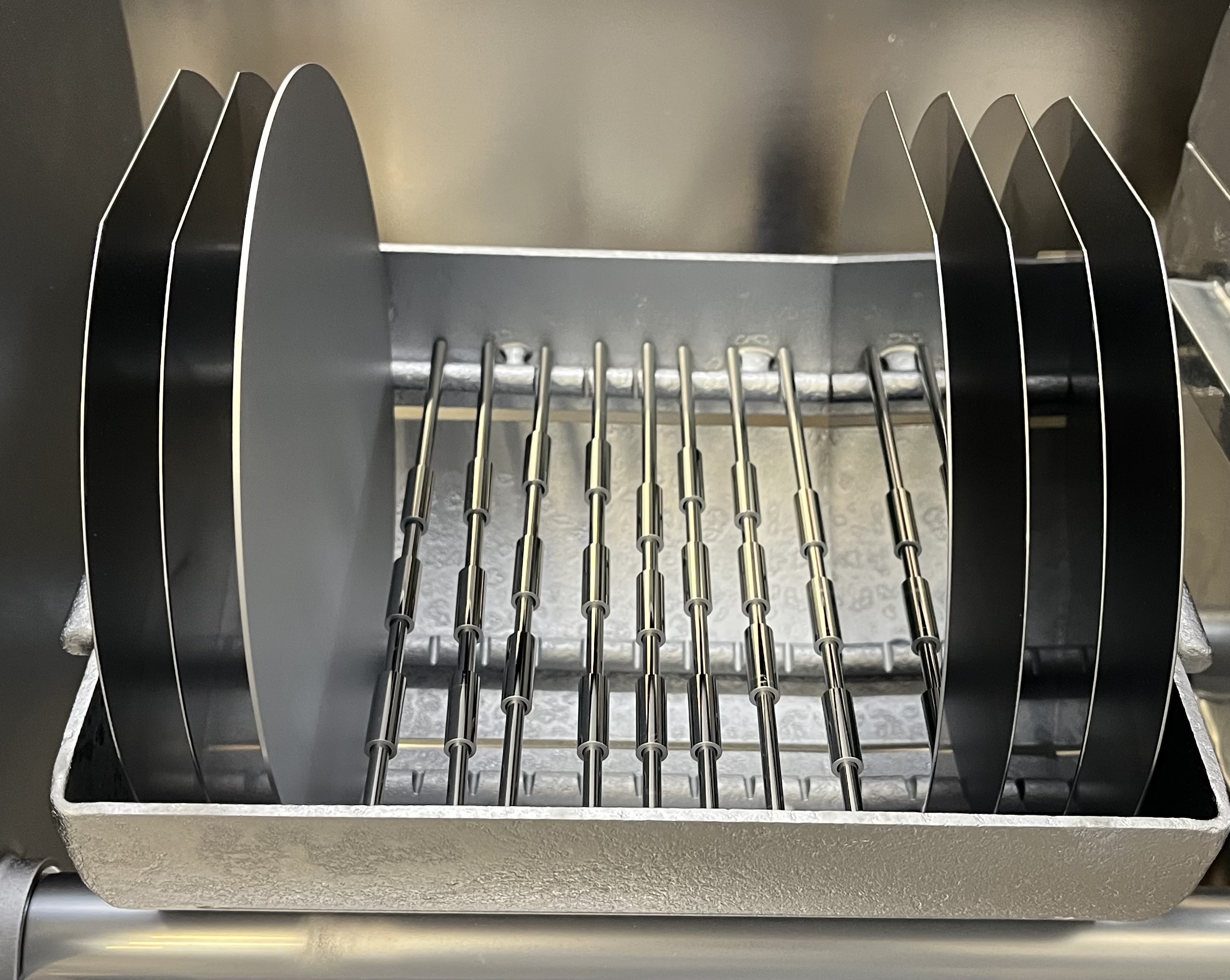}
     \caption{The setup using silica rods and supported by the the wafer ``boat'' slots, with silica spacer blanks strung on each rod.}
     \label{fig:rodsetup}
 \end{figure}

The encouraging results of these tests motivated a ``production'' run of 25 spacer/resistors. The setup used in the furnace at MBNL for the production run is shown in Figure~\ref{fig:rodsetup}, which illustrates the silica rod fixturing method. Witness wafers were included in the furnace for the production run, and the results for these wafers are given in Table~\ref{tab:produrunwitness}

\begin{table}[h] 
    \centering
    \caption{Results of production spacer/resistor run for 0.7 sccm silane/phosphine 50/50 dopant mixture flow rate (pure silane flow at 200 sccm). Shown are the sheet resistance, the aSi layer thickness, and the resistivity as provided by the witness wafer, as well as the measured room temperature part resistance.  The projected spacer/resistor resistance at 295 K for the standard dimensions of 17 mm length, 6 mm OD, 4 mm ID produced with the same doped aSi properties are given for comparison with Table~\ref{tab:production}.} \quad
    \setlength{\tabcolsep}{5pt} 
    \renewcommand{\arraystretch}{1.2}
    \rowcolors{1}{gray!15}{white}
    \begin{tabular}{|m{2.5cm}|m{2cm}|m{2cm}|m{2cm}|m{3.2cm}|}
        \hline
        \multicolumn{1}{|>{\centering\arraybackslash}m{2.5cm}}{\bfseries dopant flow [sccm]}
        & \multicolumn{1}{|>{\centering\arraybackslash}m{2cm}}{\bfseries sR [k$\Omega$/sq]}
        & \multicolumn{1}{|>{\centering\arraybackslash}m{2cm}}{\bfseries thickness [nm]}
        & \multicolumn{1}{|>{\centering\arraybackslash}m{2cm}}{$\boldsymbol{\rho}$ [$\Omega$ cm]}
        & \multicolumn{1}{|>{\centering\arraybackslash}m{3.2cm}|}{\bfseries projected spacer/resistor (295~K) [M$\Omega$]} \\
        \hline
        0.7 & $2678 \pm 435$ & $305 \pm 10$ & $81.6 \pm 13.5$ & $2.42 \pm 0.41$ \\
        \hline
    \end{tabular}
\label{tab:produrunwitness}
\end{table}

\begin{figure}[htbp]
\includegraphics[width=.54\linewidth]{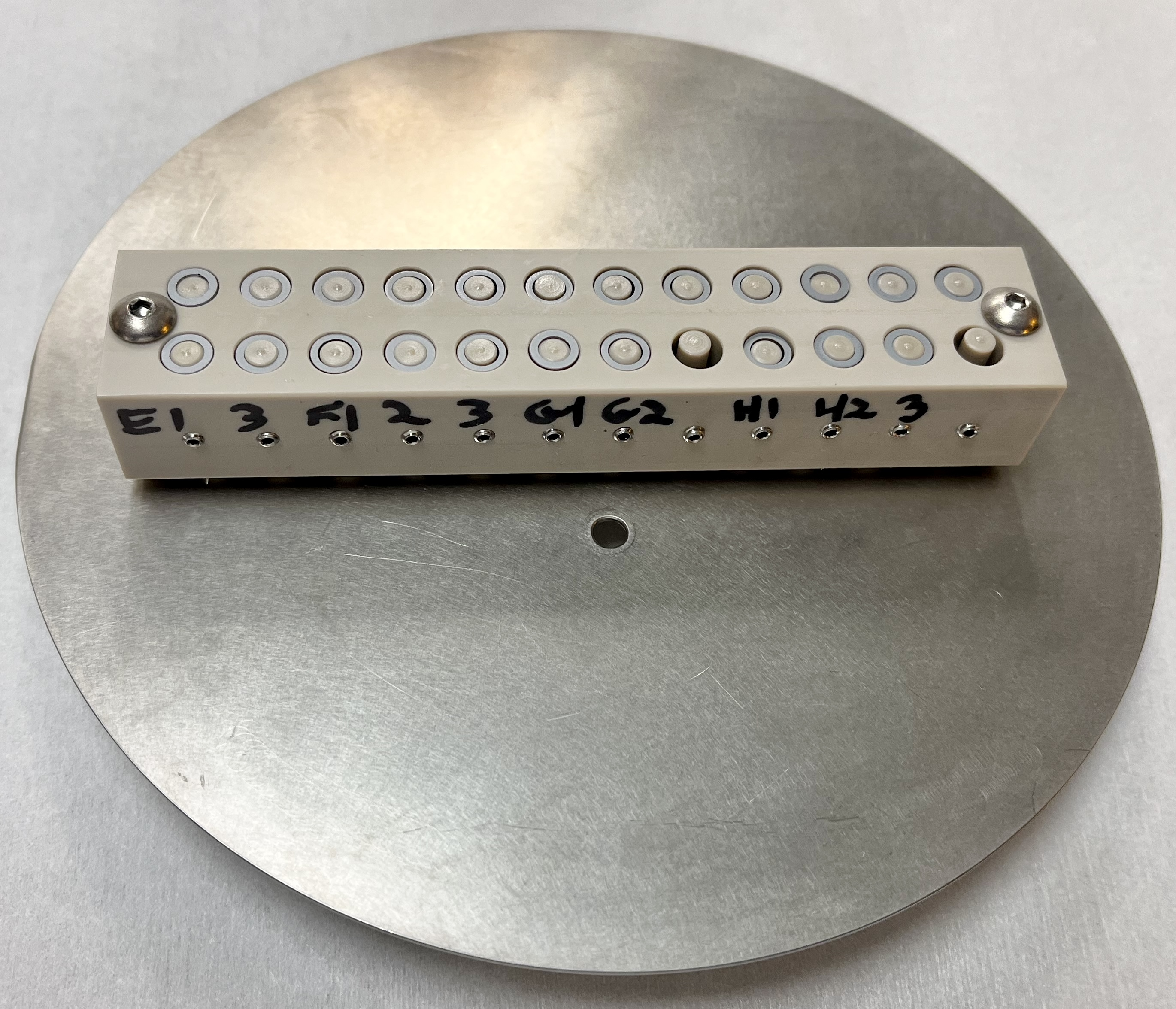}
\includegraphics[width=.46\linewidth]{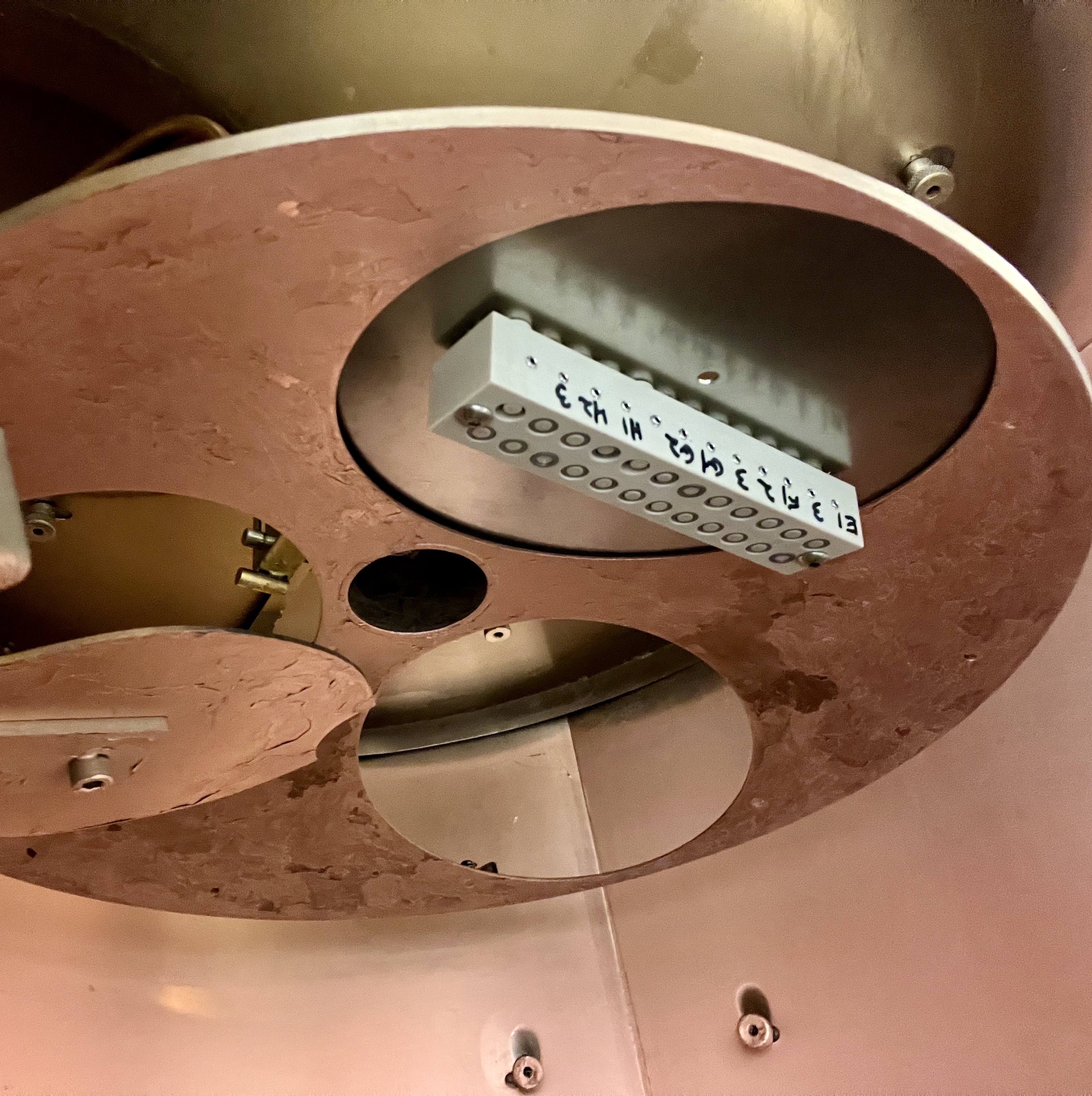}
\caption{The fixture for the metallization step (left) and its placement in the CHA (right). The metallization of each end is performed separately and the spacers are rotated 180 degrees between runs.}
\label{fig:met_photos}
\end{figure}

For the metallization step, we designed a simple fixture to hold 24 pieces, and to coat one end of each of the spacer parts per run while masking the sides of the spacer.  The fixture is then flipped over and reinstalled for a second run in the E-beam evaporator chamber to coat the opposite ends.  The fixture was machined from PEEK polymer, and included nylon tipped set screws at each spacer location to secure the part during the process.  Figure~\ref{fig:met_photos} shows the chamber setup with fixture attached to a cell normally used to coat an Si wafer and facing downwards towards the heated crucible, as well as a close up of the fixturing device.

\begin{table}[htbp]
    \centering
    \caption{The production run set of 19 spacer/resistors from MBNL.}
    \quad
    
    \renewcommand{\arraystretch}{1.2}
    \begin{tabular}{|l|c|}
    \rowcolor{gray!15}
    \hline
     \bfseries Properties & \bfseries Production Run \\
     \hline
     Mean Resistance (295~K) [M$\Omega$] & $1.63 \pm 0.01$ \\
     Standard dev (295~K) [M$\Omega$] & 0.045 \\
     Mean Resistance (165~K) [M$\Omega$] & $449 \pm 6$  \\
     Standard dev (165~K) [M$\Omega$] & 28 \\
     Mean $\Delta$TCR & $276 \pm 4$ \\ 
     \hline
    \end{tabular}
    \label{tab:production}
\end{table}

 \begin{figure}[htbp]
     \centering
     \includegraphics[width=0.5\linewidth]{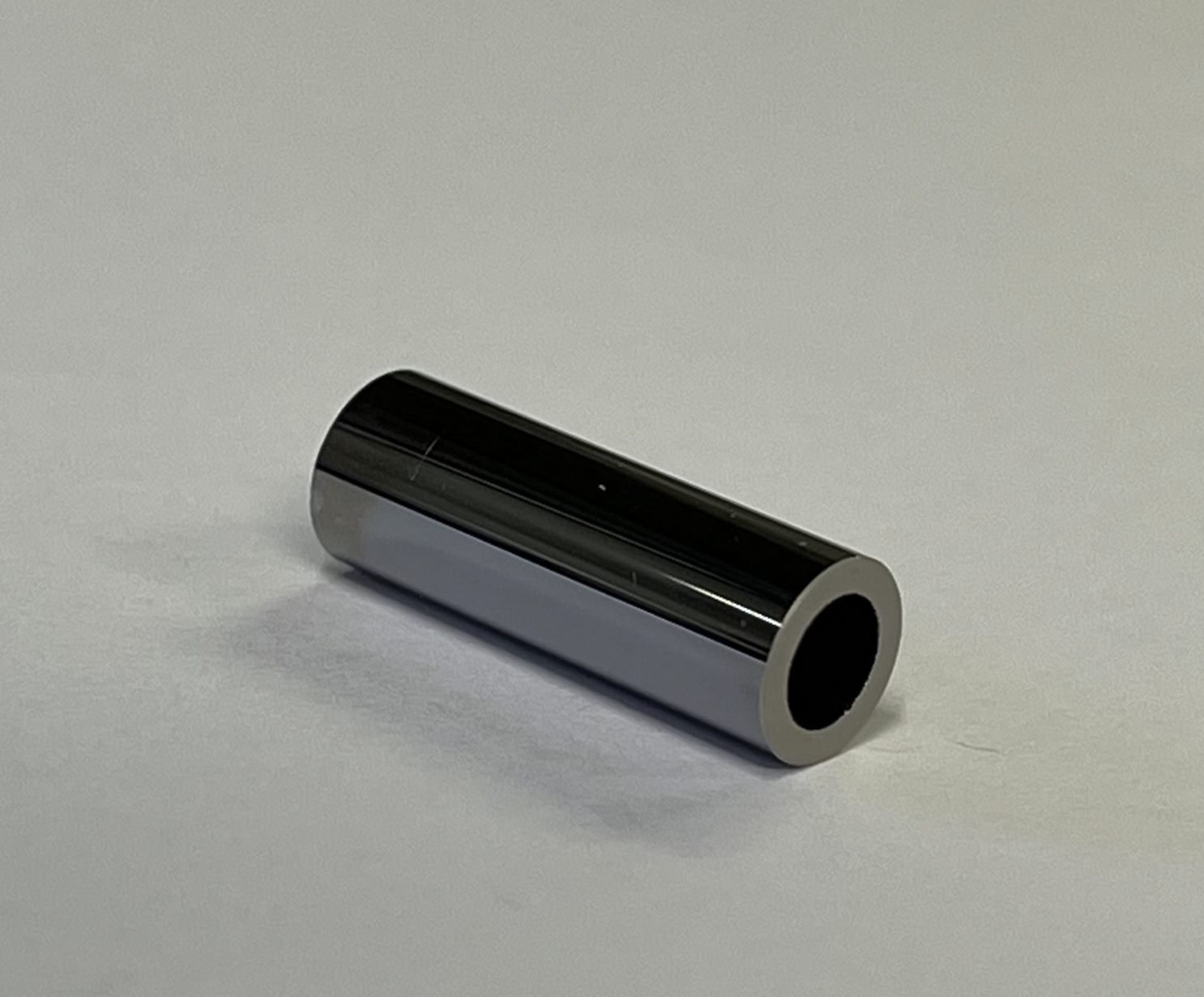}
     \caption{A 17 mm long spacer/resistor from the production run.}
     \label{fig:partphoto}
 \end{figure}

This procedure currently accommodates 20~-~25 parts per run, see Figure~\ref{fig:met_photos}.  A purpose built fixture should be good for perhaps twice that number.  The results for this production run are given in Table~\ref{tab:production}, showing the measured resistance at both room temperature and at the 165~K operating temperature.  Figure~\ref{fig:partphoto} shows one spacer/resistor from the production run.  The mirror-like aSi surface and the matte Pd-coated end are visible. There is a slight chamfer at the outer radius of the metallization, which is only 114 microns wide (see Figure~\ref{fig:SR_drawing}).

\subsection{Measurement of the thermal coefficient of resistivity (TCR)}

Measurements of the total change in resistivity from room temperature to 165 K are expressed by $\Delta \rm{TCR}$, a multiplicative factor. Table~\ref{tab:production} shows the measurement of $\Delta \rm{TCR}$ for our production run.
There is a low-significance suggestion from the data that $\Delta \rm{TCR}$ is larger for larger room temperature resistance, but here we assume that $\Delta \rm{TCR}$ and resistance are not correlated.  The relatively small spread in the 165~K resistance data suggests that ``cherry-picking'' from production runs would easily provide a selection of resistors with the desired $<20$\% variation.

\begin{figure}[htbp]
     \centering
     \includegraphics[width=0.8\linewidth]{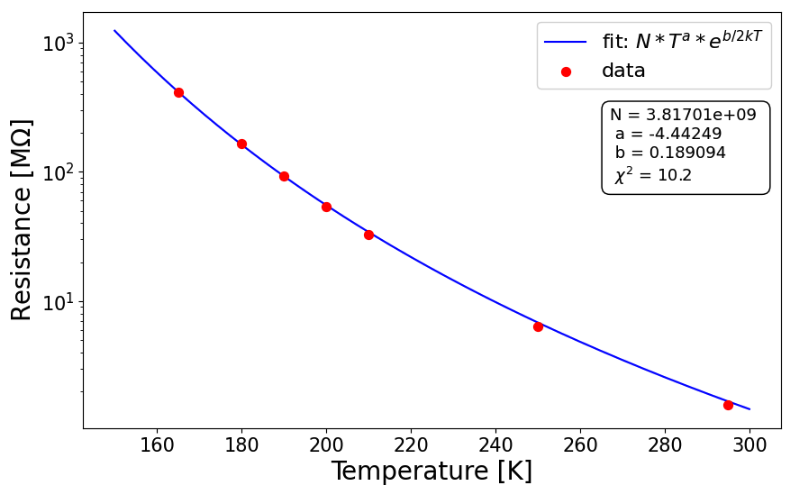}
     \caption{The fit of the resistivity data from the first production batch.}
     \label{fig:TCRfit1}
 \end{figure}

A determination of the temperature dependence of the resistivity requires a scan over a range of temperature set-points, where care is taken to reach equilibrium at each temperature value.  This measurement was done for a spacer/resistors with resistance near the mean observed room temperature value of the produced parts.  The EC was used to step the temperature from room temperature (295 K was used) to six distinct temperatures, holding steady for about 1 hour at each of the seven steps to ensure thermal equilibrium had been reached inside the testing box.  The observed resistances are plotted in Figure~\ref{fig:TCRfit1}.

The resistivity fit shown is to an ansatz  
\begin{equation}
\rho = NT^{a}e^{b/2kT}
\end{equation}
inspired by the form of the expected thermal dependence of resistivity on temperature for pure crystalline silicon, where $N$ is a normalization parameter, for which $a=-3/2$ and $b=E_g$, the 1.1~eV Si band-gap energy.

We make no attempt here to interpret the fit results for the $a$ and $b$ parameters in terms of a physical model.  The resulting temperature dependent resistivity function is a practical tool, in this case able to predict the resistivity at the 165 K operating temperature with approximately $\pm~6.7 \%$ precision, determined by our EC temperature uncertainty derived from the observed stability at the temperature set point. In any real application, the TCR fit from bench-test data would initially need to be redone for each production run of spacer/resistors.

\subsection{Photosensitivity}

As expected, silicon coatings show a degree of light sensitivity \cite{photosense}. This behavior was observed during resistivity measurements of intrinsic Si and aSi parts. Nevertheless, it does not constitute a problem since the plan is to use such components in a completely light-tight cryostat, and our measurements were performed using a light-tight box as shown Figure~\ref{fig:testing_box}. We tested two of our spacers /resistors from the production run at room temperature in room light using a DVM, with the results shown in Table~\ref{tab:phototest}.  Aside from the proviso that resistor quality-control testing is best done in dark conditions, the data do not show worrisome light sensitivity ($<$~1\% in our bench test).

\begin{table}[htbp]
    \centering
    \caption{Photosensitivity test results show small changes in the observed resistance between dark (box closed) and room light (box open) conditions.  The room light (overhead fluorescent) was measured at 6860 Lux with an Urceri MT-92H light meter.}
    \quad
    \renewcommand{\arraystretch}{1.2}
    \begin{tabular}{|c|c|c|}
    \rowcolor{gray!15}
    \hline
       \bf{production part} & \bf{R box closed [MOhm]} & \bf{R box open [MOhm]}\\
        \hline
        Spacer/Resistor 1 & 1.630 & 1.623 \\
        Spacer/Resistor 2 & 1.556 & 1.546 \\
        \hline
    \end{tabular}
    \label{tab:phototest}
\end{table}

\subsection{Comment on High Voltage testing}

To date, no spacer/resistor has been tested at a potential difference above 1 kV end-to-end. Usually, our resistance measurements with the Keithley electrometer are done at its 40~V auto setting, but tests at higher voltages up to 1 kV were made, and as expected for this modest voltage no breakdown was observed.  For the nEXO design, with a nominal operating voltage at the cathode of 50~kV, the expected potential drop across each spacer/resistor is expected to be about 860 V.  However, further HV testing to roughly 2 kV would be prudent.  In the event issues were to be observed, the standard surface features for HV insulators designed to interrupt current paths such as circumferential ribs could be added. It has not yet been necessary to investigate this mitigation scheme by simulation or testing.    

Issues regarding the Ohmic behavior of these resistors have not yet been carefully studied as the original application envisioned in the nEXO TPC does not require good linearity. It has been observed that the large TCR does introduce small non-linearity due to dissipation heat at high current operation up to 1 kV.  This effect should be mitigated by TPC cryostat temperature regulation.

\section{Radiopurity testing at PNNL}

Ultra-high radiopurity is an important requirement for most components of the nEXO detector, determined by simulation to warrant a target radiopurity at the approximately sub-10 ppt level for U and Th for the spacer/resistors. The first prototypes made at SNF were sent to our radio-assay expert collaborators within the nEXO project, based at Pacific Northwest National Laboratory (PNNL). The PNNL team employs high-sensitivity inductively coupled plasma mass spectrometry (ICPMS) techniques to measure the U and Th content of some materials at the sub-ppt level \cite{RBC2026,Harouaka2020,Arnquist2017,Arnquist2020}.  We delivered a prototype completed spacer/resistor made at SNF along with an untreated silica tube blank to PNNL for analysis.  

Concentrations are reported in ppt or pg/g for \nuc{232}{Th} and \nuc{238}{U}, and ppb or ng/g for natural K. One sample of each material was measured after full dissolution.  The uncertainty on the measurement is the analytical precision from the instrumental analysis, which is the standard deviation from three acquisitions of the sample solution on the ICP-MS. 

The results of this test were encouraging, with U and Th levels close to or below the specified level without having taken any extraordinary measures in process cleanliness besides the usual nanofabrication procedures.  These very first tests were compromised by differing sample cleaning and preparation procedures used for the completed part and the blank intended as a control measurement (see Table~\ref{tab:radiotest1}). 

This shortcoming was corrected in a test of prototypes produced at MBNL.  The second radiopurity measurement used the first prototypes produced at MBNL, and in this case care was taken to preclean all parts in the same way, including the control blank.  In addition, a partially completed part with Si coating but no metallization was provided for testing.

The silica parts were pre-cleaned with a sulfuric acid and hydrogen peroxide mixture, abbreviated as SPM. ``Piranha Etch'' is also a well-known name \cite{Piranha}.  The bath (about 18 L volume) is initially 98\% sulfuric acid $\rm {H_{2}SO_{4}}$ at 120\textdegree C. The bath is then spiked with 200 cc of 30\% hydrogen peroxide ($\rm {H_{2}O_{2}}$) immediately before use as the peroxide degrades fairly rapidly. The part is then submerged for 10 minutes followed by a cycle of 4 quick dump rinses in deionized water and then blown dry with nitrogen gas. 

\begin{table}[htbp]
    \centering
    \caption{Radiopurity test 1. *Control spacer was missing cleaning step to remove organic residue. Only parameter variation is LPCVD temperatures: Run 1: 580\textdegree C, Run 2: 550\textdegree C.}
    \quad
    
    \renewcommand{\arraystretch}{1.2}
    \begin{tabular}{|l|c|c|}
    \rowcolor{gray!15}
    \hline
       \bf{Prototype 1 SNF - 2023} & \bf{\nuc{232}{Th} [ppt]} & \bf{\nuc{238}{U} [ppt]}\\
        \hline
        *Control & 4.4 $\pm$ 0.2 & 23.2 $\pm$ 0.6 \\
        Completed run 1 & 7.1 $\pm$ 0.5 & 5.41 $\pm$ 0.1 \\
        Completed run 2 & 7.6 $\pm$ 0.3 & 10.6 $\pm$ 0.4 \\
        \hline
    \end{tabular}
    \label{tab:radiotest1}
\end{table}

\begin{table}[htbp]
    \centering
    \caption{Radiopurity test 2.}
    \quad
    
    \renewcommand{\arraystretch}{1.2}
    \begin{tabular}{|l|c|c|}
    \rowcolor{gray!15}
    \hline
        \bf{Prototype 2 MBNL - 2024} & \bf{\nuc{232}{Th} [ppt]} & \bf{\nuc{238}{U} [ppt]}\\
        \hline
        Control & 0.72 $\pm$ 0.09 & 0.22 $\pm$ 0.02 \\
        Non-metalized & 0.82 $\pm$ 0.06 & 0.54 $\pm$ 0.03 \\
        Completed & 2.00 $\pm$ 0.16 & 4.3 $\pm$ 0.2 \\
        \textbf{nEXO target} & \textbf{$<22.5$} & \textbf{$<7.4$} \\
        \hline
    \end{tabular}
    \label{tab:radiotest2}
\end{table}

The results of the second measurement are shown in Table~\ref{tab:radiotest2}.  It is evident that these parts test cleaner than our first prototypes, and below the target radiopurity for use in the detector.  It is also clear that most of the U and Th detected arise from the metallization process, and not from the silica substrate or the LPCVD Si.  

Although Heraeus silica proved satisfactory from a radiopurity point of view, the mechanical properties of sapphire are superior with respect to material toughness, resistance to fracture and suitability for compressive and tensional loading, with a larger Young’s modulus and higher tensile and compressive strengths than silica. For this reason the nEXO collaboration undertook a program to qualify sapphire material for radiopurity via a test of Saint Gobain sapphire \cite{sapphireradio}, and it is suitable for use both as a substrate for spacer/resistors and for the TPC tensioning rods. In this work we also tested LPCVD deposition integrity on sapphire, which will be briefly mentioned in our concluding section.

\section{UV Reflectivity Studies at UMASS}

Another property, particularly specific to LXe TPC applications for the spacer/resistors, is their surface reflectivity at vacuum ultraviolet (VUV) wavelengths. Energy deposition in LXe due to both energetic charged-particle or photon interactions appears as ionization of the medium, producing free electrons and excited xenon atoms (excimers) and ions.  Scintillation light is then produced by excimers de-excitation and electron/ion recombination primarily at a wavelength of 175~nm in the VUV.  The collection of both the ionization electrons and the scintillation light is essential for high-quality energy resolution and positional reconstruction.

A 4" witness wafer from our spacer/resistor run at SNF (aSi deposited at 550\textdegree C) was sent to our colleagues at UMASS who have a dedicated LXe reflectivity testing setup for a reflectivity measurement~\cite{UMASS1}.  The results they obtained are given in Table~\ref{tab:reflect} and agree within errors with the value available in the literature \cite{Pierce1972} which indicate useful reflectivity at 175~nm and in principle improves light collection. Additional data and analysis will be required before both the reflectance and the specular and diffuse components can be determined \cite{UMASS2}.  As one can see from the TPC rendering in Figure~\ref{fig:fieldcage}, for the case of the nEXO design the spacers represent a small solid angle for light reflection compared to other surfaces. However, for some applications of this aSi technique to a larger surface area, the importance of reflectivity might be more significant.

\begin{table}[htbp]
    \centering
    \caption{Total reflectivity [\%] at 175 nm at normal incidence.}
    \quad
    
    \renewcommand{\arraystretch}{1.2}
    \begin{tabular}{|c|c|}
    \rowcolor{gray!15}
    \hline
        \bf{Sample} & \bf{Total Reflectivity [\%]}\\
        \hline
        aSi(550\textdegree C) & 48 $\pm$ 15 \\
        \hline
    \end{tabular}
    \label{tab:reflect}
\end{table}

\section{Conclusions and General Applications}

This paper has described an initial R\&D program at SLAC, supported by Stanford and Berkeley nanofabrication labs, that has demonstrated the production, with low statistics, of high-radiopurity resistors suitable for use in high-voltage systems for low-background liquid xenon cryogenic experiments.  The prototype resistors met our specifications for ppt-level U and Th contamination, and demonstrated the desired resistance with satisfactory variance in a small production run. Our experience with less successful runs than the ones described in this article does reveal the need for improvements before mass production is optimized.  We identified a few straightforward engineering improvements to our fixturing, both for the furnace Si deposition and the metallization steps, that would be appropriate for a future publication.  In addition, there are outstanding general questions for further study:

\begin{itemize}

\item More extensive HV testing is needed including tests to failure.
\item Mechanical testing of the assembly of the voltage-divider within a realistic prototype field cage is needed, in particular to determine the robustness of the metallization during this process.
\item Further study of the effects of aSi light sensitivity will be necessary, particularly if the resistors are intended for use outside of strictly light-tight environments.
\item The significant and negative TCR of aSi is a unique property of these resistors, and the implications this has on their use require close attention, particularly for systems without stable temperature control.
\end{itemize}

The design of the spacer/resistor arose from the needs of the nEXO R\&D program, and its combined role as an electronic component and structural member is unusual. Although spacer/resistors in principle could be applied to other TPC designs (gas or liquid and room temperature or cryogenic), less complex versions of high-radiopurity aSi resistors are plausible.  We did not investigate the fabrication of ``flat'' aSi resistors on silica or sapphire substrate; however, we did coat a 6" sapphire wafer as a test to confirm satisfactory Si binding to sapphire and did not encounter any issue with the furnace aSi process. We also tested the aSi coated sapphire wafer down to 165~K without observing any changes in surface quality that might arise from a coefficient of thermal expansion (CTE) mismatch. 

It is possible that discrete resistors used for ultra-low background experiments such as resistors used in PMT bases or in ASIC-based amplification/digitization circuits, using the clean Heraeus 300-series Suprasil or the Saint Gobain sapphire as a substrate, might benefit from this high-radiopurity fabrication method. 

\section*{Acknowledgments}

We are grateful for the contributions of Maurice Stevens, formerly of Stanford Nanofabrication Laboratory, whose initial participation in our R\&D program was essential to get this project started, and to Dr.Mary Tang (Managing Director of SNF) for her support.  We thank Dr.William Flounders, Director of the Marvell Berkeley Nanofabrication Laboratory (MBNL). We also appreciate the insights and advice we received from Prof.Ted Kamins (Stanford).  We thank Drs.Christopher Kenney and Julie Segal (SLAC Instrumentation Division) for their expertise and support. Finally, we thank our colleagues within the nEXO Collaboration for input and discussions during this research.
\par
This research was supported by the U.S. Department of Energy, Office of Nuclear Physics, Contract DE-AC02-76SF00515 (SLAC National Accelerator Laboratory) and Contract DE-AC05-76RL01830 (Pacific Northwest National Laboratory) and by the National Science Foundation, grant NSF PHY-2111213 and U.S. Department of Energy award DE-SC0020509 (University of Massachusetts).




\bibliography{bib}

\end{document}